\documentclass{JHEP3}

\usepackage[numbers,sort&compress]{natbib}
\usepackage{amsmath}

\title{Finite-size effects for jet quenching}

\author{Suphakorn Chunlen, Kasper Peeters and Marija Zamaklar\\
Department of Mathematical Sciences\\ 
Durham University\\
South Road\\
Durham DH1 3LE\\
United Kingdom.\\
~\\
\email{suphakorn.chunlen@durham.ac.uk},\\
\email{kasper.peeters@durham.ac.uk},\\
\email{marija.zamaklar@durham.ac.uk}}

\date{20 December 2010}

\preprint{DCPT-10/79}

\abstract{We study corrections to the drag force exerted on a quark
  moving through a quark-gluon plasma of finite extent, using
  holographic methods. Interestingly we find that the leading
  correction is negative, i.e.~it reduces the magnitude of the drag
  force as compared to its value in infinite volume.}

\begin{document}

\section{Introduction}

During the last few years, much work has been devoted to studying
strongly coupled processes in the quark-gluon plasma by making use of
the gauge/gravity duality. In particular, a lot of attention has
focused on the phenomenon of jet quenching: quark-antiquark pairs
which are produced near the boundary of the quark-gluon plasma do not
lead to two back-to-back jets, but rather give rise to only one
observed jet. A simple qualitative explanation of this phenomenon lies
in the fact that the quark which needs to move through the plasma
before it can escape loses a lot of its initial energy due to
interaction with the medium.  While this is a qualitatively and
intuitively simple explanation, it is hard to obtain a quantitatively
correct answer from it. Starting with the work of
\cite{Gubser:2006bz,Herzog:2006gh}, the ultra-relativistic quenched
quark produced inside the strongly coupled quark-gluon plasma was
modelled holographically with an open string ending on the boundary of
AdS space, hanging deep in the interior of an AdS black hole. This
initial work, as well as subsequent generalisations to other
dimensions, to presence of chemical potentials or higher derivative
corrections, all focused on the study of quark motion in an
infinite-volume medium.

A realistic quark-antiquark pair produced in colliders, however, 
propagates in a plasma of finite extent. The question then arises
whether this has any influence on the jet quenching parameter as
computed in an infinite volume system. In this paper we initiate the
study of finite-size corrections to the quark motion using the
gauge/gravity correspondence.

To achieve this we will study, in a holographically dual picture, a
string which is moving in the background of a black hole in global AdS
space, as opposed to the planar AdS black holes in the Poincar\'e
patch which were studied so far.  Black holes in the Poincar\'e patch
have horizons which are non-compact and planar, and as such they are
dual to a quark-gluon plasma of infinite extent. In contrast, black
holes in global AdS space have spherical horizons and are dual to a
quark-gluon plasma on a three-sphere.

The only scales in the finite-temperature super-Yang-Mills theory on a
sphere are the temperature~$T$ and the radius of the three sphere~$L$,
so that the only dimensionless quantity one can form is the product
$TL$. Therefore, all results, including the dimensionless drag force
which we will obtain, will be functions of this quantity. Although
global AdS and planar black holes are genuinely different
gravitational configurations, there is a limit in which a global black
hole reduces to the planar one.  In the limit of a large global AdS
black hole, when the size of the horizon is much larger than the AdS
scale, $ \rho_H \gg L$, the global AdS black hole reduces to the
planar one.  This limit translates to the condition $TL\gg 1$, which
can be interpreted either as the large-volume limit at fixed
temperature, or as the high-temperature limit at fixed sphere
size. Hence, in the large black hole limit $\rho_H \gg L$ we expect to
recover the result of the infinite volume system (the planar black
hole), amended by a tower of finite-size corrections.

Indeed this is what we will find in this note. Interestingly, the
leading correction which we find is negative, i.e.~it reduces the
magnitude of the drag force. This is in contrast to what one usually
encounters in situations with large objects moving in a
non-relativistic fashion through fluids in finite-size containers. In
these systems one usually finds that the drag force is increased by
the presence of walls (see e.g.~\cite{Chang:1961a,Song:2009a}). The
basic intuitive reason for this behaviour is that the walls impose an
additional friction force on the fluid, creating a contra-flow which
further obstructs propagation of the object.

In our setup the situation is a bit different, since the quark-gluon
plasma is placed on a sphere, i.e.~in a container without
boundaries. However, in our case finiteness of space is imposed
through the periodicity condition which any excitation in the fluid
has to satisfy. Therefore, not any arbitrary excitation which is
present in infinite volume can be excited in this system. This is a
possible reason why the force is reduced with respect to the one in
infinite volume. It would be very interesting to test finite-size
corrections for the quark-gluon plasma in other systems, hopefully
those which do have boundaries, and use holography to check which of
the features observed here persist. One possible system which one
could consider is the numerical solution of \cite{Aharony:2005bm}
which describes a classically stable finite energy black hole
localised in the infrared part of the geometry.
\smallskip

{\it Note added:} When this paper was completed we became aware of the
work \cite{Atmaja:2010hd} which has some overlap with our results in
section~2.

\section{Dragged string in a global AdS black hole}

The system we will study is the Schwarzschild black hole in
\emph{global} $\text{AdS}_5$ space, in contrast to the Poincar\'e patch
studied before \cite{Gubser:2006bz,Herzog:2006gh}.  The metric of the
global $\text{AdS}_5$ black hole is given by
\begin{equation}
\begin{aligned}
{\rm d}s^2 &= - h(\rho) {\rm d}\tau^2 + \frac{1}{h(\rho)} {\rm
  d}\rho^2
+ \rho^2 \big({\rm d} \theta^2 + \cos^2\theta ({\rm d}\phi^2 + \cos^2\phi {\rm d}\chi^2)\big)\,, \\[1ex]
\rho_0^2 &= \frac{8 GM}{3 \pi} \, , \quad \quad h(\rho) = 1 - \frac{\rho_0^2}{\rho^2} + \frac{\rho^2}{L^2} \,.
\end{aligned} 
\end{equation}
Here $G$ is the Newton constant, $M$ is the mass of the black hole and
$L$ is the radius of the AdS space. The boundary of the AdS space is
at $\rho \rightarrow\infty$, while in the interior there is a black
hole horizon at position $\rho_H$,
\begin{equation}
\label{variousa}
\rho_H = L \sqrt{\frac{\sqrt{1 + 4 \rho_0^2/L^2} -1}{2}} \, ,  \quad \quad 
 \rho_0 = \rho_H \sqrt{1 + \frac{\rho_H^2}{L^2}} \, .
\end{equation}
The temperature of the black hole is given by~\cite{Gibbons:2004ai}
\begin{equation}
\label{temperature}
T = \frac{\rho_H}{\pi L^2} \left( 1 + \frac{L^2}{2 \rho_H^2}\right)  \, .
\end{equation}
As the temperature is decreased, on encounters a first-order phase
transition at the critical temperature $T_{\text{HP}} = 3/(2\pi L)$
below which pure $\text{AdS}_5$ space (with periodic time circle) is
preferred over $\text{AdS}_5$-Schwarzschild
\cite{Witten:1998zw,Hawking:1983dh}. At this temperature the boundary
theory undergoes a ``deconfinement/confinement'' phase transition, in
the sense that the free energy of the system is of order $N_c^2$ above
the temperature $T_{\text{HP}}$ and of the order one below it. We also note
that the only scales in the super-Yang-Mills theory are the AdS radius
$L$ (which is the same as the size of the boundary sphere) and the
temperature $T$, so that the only dimensionless and physically
relevant parameter is $TL$.  Hence the limit of large $TL$ can
equivalently be interpreted either as the high-temperature limit at
fixed volume or the large-volume limit at fixed temperature. We will
be working at fixed temperature and interpret $TL\rightarrow$ as the
limit of large volume.

The $\text{AdS}_5$ Schwarzschild black hole should be compared with the planar
AdS black hole,
\begin{equation}
\label{planarspace}
{\rm d}s^2 = \frac{r^2}{L^2} \left(-\left(1 - \frac{r_H^4}{r^4}\right) {\rm d}t^2 + {\rm d}\vec{x}^2\right) + \frac{L^2}{r^2} \frac{{\rm d}r^2}{1 - \frac{r_H^4}{r^4}} \,.
\end{equation}
the boundary of which is ${\mathbb R}^3 \times S^1$ with $S^1$ being
the time circle. We will later use the fact that the global
$\text{AdS}_5$ Schwarzschild black hole reduces to the planar one in
the limit of a large black hole.

\medskip 

In order to describe the dragged string in the global AdS black hole,
we use world-sheet coordinates aligned with $\tau$ and $\rho$, and an
embedding given by
\begin{equation}
\theta = \omega \tau + f(\rho) \, .
\end{equation}
With this ansatz the action for the string becomes
\begin{equation}
S = \int\sqrt{1 + \rho^2 f'{}^{2} h(\rho)  - \frac{\rho^2 \omega^2}{h(\rho)}} \, .
\end{equation}
Next, we introduce the ``conserved momentum'' $\pi_f = \partial S/\partial
  f'$, conjugate to $f$, in terms of which $f'$ is
\begin{equation}
\label{fprimess}
f'(\rho) = \frac{\pi_f}{\rho\, h(\rho)} \sqrt{\frac{\rho^2 \omega^2 - h}{\pi_f^2 - 
\rho^2 h}}\,.
\end{equation}
Substituting this back into the action we get
\begin{equation}
\label{actionss}
S = \rho  \sqrt{\frac{\rho^2 \omega^2 - h}{\pi_f^2 - \rho^2 h}}\,.
\end{equation}
As in the planar case, we fix $\pi_f$ from the requirement that $f'$
and the action are real functions. In other words, we require that
when the numerator inside the square root of \eqref{fprimess} and
\eqref{actionss} changes sign, the denominator changes sign as
well. The position $\rho=\rho_*$ at which this happens is determined
by
\begin{equation}
  \rho_*^2 \omega^2 - h(\rho_*) = 0 \, , \quad \quad  \pi^2_f =
  \rho^2_* h(\rho_*)
 = \rho^4_* \omega^2\,.
\end{equation}
This can be solved to obtain
\begin{equation}
\label{piiiffss}
\pi_f = \omega L^2 \frac{\displaystyle 
-1 \pm  \sqrt{1 + 4 \frac{\rho_0^2}{L^2} (1- \omega^2 L^2)}}{2
  (1 - \omega^2 L^2 )} \, .
\end{equation}
We note that we should take the plus sign in this expression, since in
the pure AdS background (i.e.~\mbox{$\rho_0=0$}) there is no
dissipation and the drag force should be zero.

Obtaining the shape of the dragged string in the global AdS space is
more complicated than in the planar case, since it requires
integration of \eqref{fprimess} using the expression
\eqref{piiiffss}. This can be done numerically, but we do not need
this information in order to extract the drag force itself.

In order to compute the loss of the energy of the dragged string, we
need to evaluate the flow of the momentum down the string, towards the
horizon. Following \cite{Gubser:2006bz} we thus need to compute
\begin{equation} 
\frac{{\rm d}p_\theta}{{\rm d}t} =\sqrt{g} P^\rho{}_\theta \, , \quad
 P^\alpha{}_\mu \equiv - \frac{1}{2 \pi \alpha'}  G_{\mu \nu} g^{\alpha\beta} \partial_\beta X^\alpha\,,
\end{equation}
where $P^\alpha_\mu$ is the conserved world-sheet current of the
space-time energy momentum, and $g_{\alpha\beta}$ and $G_{\mu\nu}$ are
the induced metric on the world-sheet and the target space metric
respectively. We find
\begin{equation}
\label{dpdt}
\frac{{\rm d}p_\theta}{{\rm d}t} = -\frac{1}{2 \pi \alpha'}\pi_f\,,
\end{equation}
with $\pi_f$ given by \eqref{piiiffss}.

\section{Expansion near the planar AdS black hole}

In order to make contact with results for the drag force obtained in
the planar case, we now wish to expand the result~\eqref{dpdt} given
\eqref{piiiffss}. In the case of non-relativistic fluids with small
Reynolds number, the drag force acting on a moving object can
typically be written as~\cite{Chang:1961a,Song:2009a} $\vec{F} =
f(\eta, d)\, \vec{v}$ where the function $f(\eta,d)$ depends on the
viscous properties of the fluid (through the viscosity constant
$\eta$) and the size of the object $d$. This expression is valid only
if the size of the container in which the particle moves is
infinite. In the case of a container with finite size $D$, one expects
that $f$ will also depend on the dimensionless ratio $d/D$. Typically
the function $f$ is then hard to compute and it is extracted only
experimentally. We now proceed to analyse these type of corrections to
the drag force for our system.

Before continuing, let us make a comparison of \eqref{piiiffss} with
the analogue quantity in the planar case. If the string is moving
along the $x$-direction in the planar background~\eqref{planarspace},
with embedding given by $x= vt + \xi(r)$, then the momentum conjugate
to $\xi$ is \cite{Gubser:2006bz}
\begin{equation}
\label{piiixiii}
\pi_\xi = \frac{r_H^2}{L^2} \frac{v}{\sqrt{1 - v^2}} \, .
\end{equation}
The drag force for this planar case is
\begin{equation}
\begin{aligned}
\label{fromAdStoSYM}
\frac{{\rm d}p_x}{{\rm d}t} = -\frac{1}{2 \pi \alpha'} \pi_\xi &=
-\frac{r_H^2/L^2}{2 \pi \alpha'} \frac{v}{\sqrt{1 - v^2}} 
 = - \frac{\pi \sqrt{\lambda}}{2}\tilde{T}^2 \frac{v}{\sqrt{1 - v^2}} \,, \\[1ex]
\text{with}~~ L^4 &= \lambda \alpha'^2 \, , \quad \quad  \tilde{T} = \frac{r_H}{\pi L^2} \, ,
\end{aligned}
\end{equation}
and the 't Hooft coupling $\lambda = g_{\text{YM}}^2 N$.

In order to compare this to the global AdS black hole result, we take
the limit of a large black hole
\begin{equation}
\rho_H \gg L \,.
\end{equation}
In this limit the $S^3$ at the boundary becomes a plane, the
coordinate $L \theta$ becomes $x$ and hence the momentum $\pi_f/L$
becomes $\pi_\xi$. The velocities are related by $\omega L \rightarrow
v$. Keeping $v$ fixed we can expand \eqref{piiiffss} in powers of
e.g.~$L/\rho_0$, to get
\begin{equation}
\frac{\pi_f}{L} = \frac{-\omega L}{2(1- \omega^2 L^2)}
+ \frac{\omega L}{\sqrt{1- \omega^2 L^2}} \frac{\rho_0}{L} \bigg( 1 + \frac{1}{8} \frac{L^2}{\rho_0^2} \frac{1}{1 - \omega^2 L^2} + \cdots \bigg)\,.
\end{equation}
In order to express the drag force in terms of gauge theory
quantities, we then need the temperature expressed in terms of the
parameter $\rho_0$ appearing in the metric, using \eqref{temperature}.
It is also convenient to introduce the dimensionless force
$\mathcal{F} \equiv F L^2$, for which we finally find
\begin{equation}
\label{Fdimensionless}
\mathcal{F} =  - \frac{v\sqrt{\lambda}}{2\pi\sqrt{1-v^2}}\Big[
   \pi^2 (TL)^2 
 - \frac{1+\sqrt{1-v^2}}{2\sqrt{1-v^2}}
 + \frac{1}{8\pi^2}\frac{3v^2 - 2}{(1-v^2) (TL)^2}
+ \ldots\Big]\,.
\end{equation}
The leading term agrees with the flat result \eqref{fromAdStoSYM} and
there is a whole tower of corrections on top. We discuss these
results in the next section.

\section{Discussion}

There are several comments which we want to make about the result
(\ref{Fdimensionless}). We see that all corrections to the
leading, infinite-volume results are given in terms of the
dimensionless quantity $TL$. As we have explained in the introduction,
in conformal field theory on a sphere and at finite temperature, this
is the only dimensionless quantity available. Considering now that the
estimate for the mean free path is given by \cite{Son:2007vk}
$l_{\text{mfp}} \sim 1/\lambda^2 T$, one can interpret the
dimensionless ratio $TL$ as the ratio of the size of the plasma
container $L$ and an ``effective size'' of the quark, set by the mean
free path $\sim 1/T$. Hence the form of the corrections is as expected
on general grounds. However, while the fact that the leading-order
force~\cite{Gubser:2006bz} is proportional to $T^2$ is fixed by
dimensional analysis, the functional dependence on the velocity in the
various corrections is not.

We also note that the sign of the first correction to the infinite
volume result of \cite{Gubser:2006bz} is always opposite from the
leading term, i.e.~the magnitude of the drag force seems to be
decreased due to the finite-volume effects. This is a quite unusual
behaviour as compared to most non-relativistic Newtonian fluids.

It is conceivable that an explanation of this interesting feature lies
in the fact that our system is placed in a container without boundary,
namely the three-sphere. This basically requires that all
hydrodynamical modes with which the quarks can interact, and which it
can dump energy into, are periodic. Therefore, fewer available
hydrodynamical modes could lead to a smaller drag force, somewhat
similar to the discussion in \cite{Gubser:2009qf}. It would be very
interesting to check in other duals of plasmas in finite size systems
if such a behaviour persists.
\bigskip

\setlength{\itemsep}{5pt}
\bibliographystyle{kasper}
\begingroup\raggedright\endgroup

\end{document}